\documentclass[twocolumn]{aastex6}

\usepackage{natbib}
\usepackage{color}
\usepackage{mathtools}
\usepackage{hyperref} 

\def	\cm		{\,{\rm {cm}}}
\def	\K		{\,{\rm K}}
\def	\g		{\,{\rm {g}}}
\def	\mum	{\,{\mu \rm{m}}}

\def \bea {\begin{eqnarray}}
\def \ena {\end{eqnarray}}

\def	\B	{{\rm B}}

\def	\cm	{\,{\rm cm}}

\def	\erg	{\,{\rm erg}}

\def	\g	{\,{\rm g}}
\def	\gas	{\,{\rm gas}}

\def	\H	{{\rm H}}

\def	\IR	{{\rm IR}}

\def	\s	{\,{\rm s}}

\def	\rad	{{\rm rad}}

\def    \gas     	{{\rm gas}}

\begin{document}
\shorttitle{Dynamical Constraint of Dust Model}
\shortauthors{Thiem Hoang}
\title{A dynamical constraint on interstellar dust models from radiative torque disruption}

\author{Thiem Hoang\altaffilmark{1,2}}
\affil{$^1$ Korea Astronomy and Space Science Institute, Daejeon 34055, Republic of Korea; \href{mailto:thiemhoang@kasi.re.kr}{thiemhoang@kasi.re.kr}}
\affil{$^2$ Korea University of Science and Technology, 217 Gajeong-ro, Yuseong-gu, Daejeon, 34113, Republic of Korea}

\begin{abstract}
Interstellar dust is an essential component of the interstellar medium (ISM) and plays critical roles in astrophysics. Achieving an accurate model of interstellar dust is therefore of great importance. Interstellar dust models are usually built based on observational constraints such as starlight extinction and polarization, but dynamical constraints such as grain rotation are not considered. In this paper, we show that a newly discovered effect by Hoang et al., so-called RAdiative Torque Disruption (RATD), can act as an important dynamical constraint for dust models. Using this dynamical constraint, we derive the maximum size of grains that survive in the ISM for different dust models, including contact binary, composite, silicate-core and amorphous carbon mantle, and compact grain model for the different radiation fields. We find that the different dust models have different maximum size due to their different tensile strengths, and the largest maximum size corresponds to compact grains with highest tensile strength. We show that the composite grain model cannot be ruled out if constituent particles are very small with radius $a_{p}\le$ 25 nm, but large composite grains would be destroyed if the particles are large with $a_{p}\ge 50$ nm. We suggest that grain internal structures can be constrained with observations using the dynamical RATD constraint for strong radiation fields such as supernova, nova, or star-forming regions. Finally, our obtained results suggest that micron-sized grains perhaps have compact/core-mantle structures or have composite structures but located in regions with slightly higher gas density and weaker radiation intensity than the average ISM.
 
\end{abstract}
\keywords{ISM: dust-extinction, ISM: general, radiation: dynamics, polarization, magnetic fields}

\section{Introduction}\label{sec:intro}
Interstellar dust is an essential component of the interstellar medium (ISM) and plays critical roles in astrophysics, including gas heating, star and planet formation, and grain-surface chemistry (see \citealt{2003ARA&A..41..241D} for a review). Dust polarization induced by grain alignment allows us to measure magnetic fields in various astrophysical environments (see \citealt{Andersson:2015bq} and \citealt{LAH15} for recent reviews). Observations of starlight extinction and polarization combined with spectroscopic analysis reveal that interstellar dust includes two major components, silicate and carbonaceous materials and have different sizes and non-spherical shapes. As a result, constructing a standard model for interstellar dust is a fundamental scientific task, which has a broad application in many sub-fields of astrophysics.  

An interstellar dust model must include three ingredients: grain composition, grain geometry (shape and internal structure), and grain size distribution (see \citealt{2009ASPC..414..453D} for a review). Currently, there are three popular models of interstellar dust. The first dust model constructed from the observed wavelength-dependence extinction of starlight assumes two separate components of silicate and carbonaceous materials (\citealt{Mathis:1977p3072}; \citealt{1984ApJ...285...89D}). Later, an additional component of ultrasmall carbonaceous grains, namely polycyclic aromatic carbons (PAHs), is introduced, constituing a PAH-silicate-graphite model (\citealt{2001ApJ...554..778L}; \citealt{2007ApJ...657..810D}). The second, composite grain model is introduced by \cite{1989ApJ...341..808M} in which the grain consists of both silicate and carbonaceous particles loosely bounded together by adhesion forces. The third, core-mantle model comprises a silicate core and amorphous carbonaceous mantle (\citealt{1996A&A...309..258G}; \citealt{2013A&A...558A..62J}). Although the PAH-silicate-graphite model is widely used in astrophysics, a remaining question raised in \cite{2003ARA&A..41..241D} is "Are there really separate populations of carbonaceous grains and silicate grains? If so, how does grain growth in the ISM maintain these separate populations?"

An accurate model for interstellar dust is of great importance for developing an accurate foreground polarization model, which is urgently needed for precise detection of Cosmic Microwave Background (CMB) B-mode signal (see \citealt{Kamionkowski:2016bb} for a review). Currently, all polarized foreground models assume two distinct dust populations (\citealt{2009ApJ...696....1D}; \citealt{Guillet:2017hg}; \citealt{2018ApJ...853..127H}). Yet, it is known that a model with mixed silicate and amorphous carbon materials would produce different polarization spectra, resulting in the frequency degeneracy (see e.g., \citealt{Guillet:2017hg}).

Silicate and carbonaceous grains are formed from distinct environments, with the former dust being formed in the envelope of O-rich Asymptotic Giant Branch (AGB) stars and the later one being formed in the envelope of C-rich AGB stars. Intuitively, it is hard to believe that these two populations are completely separate in the ISM because mixing can naturally occur during the grain growth process in the ISM which is thought to be a dominant source of interstellar dust (\citealt{2009ASPC..414..453D}).  

Polarimetric observations are particularly useful to differentiate the dust models. Observations found that the 3.4 $\mum$ C-H absorption feature is negligibly polarized (\citealt{1999ApJ...512..224A}; \citealt{2006ApJ...651..268C}), whereas the 9.7$\mum$ Si-O feature is strongly polarized (\citealt{1988MNRAS.230..629A}) for the light of sights toward the Galactic Center. It is suggested that carbonaceous grains must be a separate component and these grains should be not aligned \cite{2006ApJ...651..268C}. \cite{2002ApJ...577..789L} originally argued that the lack of 3.4 $\mum$ polarization is insufficient to reject core-mantle model. A detailed study by \cite{Li:2014dk} shows that the polarization of $3.4\mum$ feature produced by the core-mantle model still exceeds the observational upper limit, which supports the original idea of two separate dust populations \citep{2006ApJ...651..268C}. Theoretically, if carbonaceous grains are separate, they are not expected to be efficiently aligned (see section 8.2 in \citealt{2016ApJ...831..159H}; recent reviews by \citealt{Andersson:2015bq}; \citealt{LAH15}). If silicate and carbonaceous components are separate as suggested by the non-detection of 3.4 $\mum$ C-H polarization, then, what physical mechanism prevents such a mixed grain model to exist in the ISM? 

A standard procedure in constructing a dust model based on observational constraints is varying the grain size distribution to achieve the best-fit model (see e.g., \citealt{Mathis:1977p3072}; \citealt{1994ApJ...422..164K}; \citealt{1997A&A...323..566L}; \citealt{2001ApJ...548..296W}; \citealt{2004ApJS..152..211Z}; \citealt{2009ApJ...696....1D}; \citealt{2017ApJ...836...13H}). While the lower cutoff of the size distribution is physically determined by thermal sublimation of nanoparticles (\citealt{1989ApJ...345..230G}), the upper cutoff is purely constrained by observational data. The latter issue is no longer valid in light of the recent progress in dust astrophysics.

Indeed, \cite{2018arXiv181005557H} discovered that large dust grains can be completely disrupted when the centrifugal stress induced by suprathermal rotation driven by radiative torques (RATs; \citealt{1996ApJ...470..551D}; \citealt{Hoang:2008gb}; \citealt{2009ApJ...695.1457H}) exceeds the maximum tensile stress (i.e., tensile strength) of dust grains.\footnote{The rotational disruption can also work for nanoparticles that are spun-up to suprathermal rotation by supersonic neutral drift in C-shocks \citep{Hoang:2018uw}.} Because the efficiency of radiative torque disruption (RATD) mechanism depends on the radiation intensity and the grain tensile strength which is determined by grain internal structures (i.e., grain model), the upper cutoff of the grain size distribution cannot be a free parameter. It should be {\it dynamically} related to the internal structure and ambient conditions. The goal of this paper is to introduce a {\it dynamical constraint} for dust models using the RATD mechanism and to explore its potential application for probing grain internal structures with observations.

The structure of the paper is as follows. In Section \ref{sec:grainmod} we will describe three popular dust models and calculate their tensile strengths. Section \ref{sec:rot} is devoted to calculate rotation rate of grains by radiative torques. We will derive the critical grain size for rotational disruption for the different dust models in Section \ref{sec:disr}. In Section \ref{sec:disc} we discuss the importance of the introduced dynamical constraint for the different dust models and the possibility of probing internal structures of dust grains with observations combined with RATD effect. A summary of our main findings is given in Section \ref{sec:summ}.

\section{Dust Models and Tensile Strength}\label{sec:grainmod}
\subsection{Contact binary model}
\begin{figure*}
\centering
\includegraphics[scale=0.3]{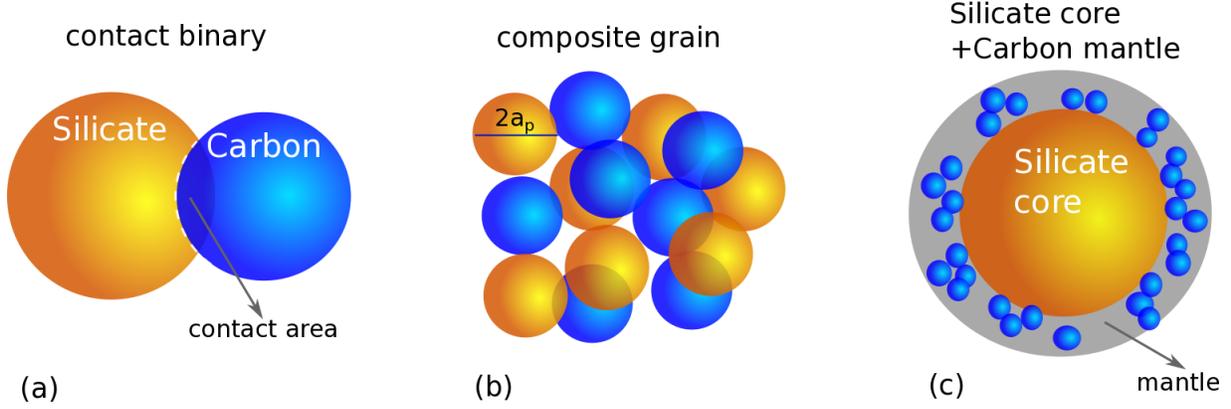}
\caption{Schematic illustration of three different grain models that form from mixing of silicate and carbonaceous grains: (a) Contact binary grain model of a spherical silicate core in contact with a carbonaceous particle, (b) A composite grain consisting of many individual particles of silicate and carbonaceous materials, (c) A silicate core and carbon icy mantle model.}
\label{fig:grain_mod}
\end{figure*}

We first consider a possible scenario in which a compact silicate particle of radius $R_{1}$ collides with a carbonaceous particle of radius $R_{2}$ in the ISM to form a contact binary grain. Upon collision, two particles make a contact area, as shown in Figure \ref{fig:grain_mod} (model a).

The physics of contact solids is well studied in literature \citep{1971RSPSA.324..301J}. The underlying physics is as follows. When two solid spheres are in contact, van der Waals forces tend to pull two spheres together. At the same time, repulsive force between nuclei act to push them. The equilibrium is established when the attractive force is equal to repulsive force. As a result, a common volume of circular area with radius $a_{0}$ is established, with the value $a_{0}$ dependeing on materials (\citealt{Chokshi:1993p3420}; \citealt{1997ApJ...480..647D}). The adhesive force due to the contact is calculated using the model of \cite{1971RSPSA.324..301J} (i.e., JKR model; see also \citealt{1999PhRvL..83.3328H}):
\bea
F_{\rm JKR}=3\pi R \gamma,
\ena
where $\gamma$ is the surface energy per unit area of the material, and $R=R_{1}R_{2}/(R_{1}+R_{2})$ is the sphere radii. The value of $\gamma$ is given by
\bea
\gamma = \gamma_{1}+\gamma_{2}-2\gamma_{12},
\ena
where $\gamma_{12}$ is the interface energy. For similar materials, $\gamma_{1}=\gamma_{2}$ and $\gamma_{12}=0$. The surface energy value varies from $\gamma=10-25\erg\cm^{-2}$ \citep{1999PhRvL..83.3328H}. 


The interaction force can be rewritten as
\bea
F_{\rm JKR}\simeq 10^{-3}\gamma_{1}R_{-5} {~\rm dyne},\label{eq:FJKR}
\ena
where $\gamma_{1}=(\gamma/10 {\rm dyn}\cm^{-2})$, $R_{-5}=R/10^{-5}\cm$. The force $F_{\rm JKR}$ is the same as the pull-off force $F_{c}$ that is required to pull two spheres apart.

Let $R_{2}=s R_{1}$ with $s\le 1$. Let $a$ be the effective grain size, which is defined as the radius of the equivalent sphere of the same volume. Then, one obtains 
\bea
a^{3}\sim R_{1}^{3}+R_{2}^{3}=R_{1}^{3}(1+s^{3}),\label{eq:aeff_binary}
\ena
where the contact area is small compared to the grain size.

\subsection{Composite grain model}
We now consider a composite grain model as proposed by \cite{1989ApJ...341..808M}. This composite model relies on the fact that upon entering the ISM, original silicate and carbonaceous grains are shattered (e.g., by shocks) into small fragments. The subsequent collisions of these fragments reform interstellar composite grains. Following \cite{1989ApJ...341..808M}, individual particles are assumed to be compact and spherical of radius $a_{p}$.\footnote{In realistic conditions, one cannot have the same size $a_{p}$ because of grain-grain collisions of different sizes. For the sake of simplicity without loosing the underlying physics, we approximate the particles to have an average size.} Particles can be of silicate or carbonaceous materials. 
Let $P$ be the porosity which is defined such that the mass density of the porous grain is $\rho=\rho_{0}(1-P)$ with $\rho_{0}$ being the mass density of fully compact grain. The value $P=0.2$ indicates an empty volume of $20\%$. 

To calculate the tensile strength of a composite grain, we follow the approach in \cite{1995A&A...295L..35G} where the particle is assumed to have a icy mantle, which is plausible because grain coagulation by grain-grain collisions is expected in cold environments such as outflows.

Let $\bar{E}$ be the mean intermolecular interaction energy at the contact surface between two particles and $h$ be the mean intermolecular distance at the contact surface. Let $\beta$ be the mean number of contact points per particle between 1-10. The volume of interaction region is $V_{\rm int}=(2ha_{p}^{2})$. Following \cite{1995A&A...295L..35G}, one can estimate the tensile strength as given by the volume density of interaction energy
\bea
S_{\max}=3\beta(1-P)\frac{\bar{E}}{2ha_{p}^{2}}.\label{eq:Smax_comp1}
\ena

We can write $\bar{E}=\alpha 10^{-3}$ eV where $\alpha$ is the coefficient of order of unity when the interaction between contact particles is only van der Waals forces. The contribution of chemical bonds between ice molecules can increase the value of $\alpha$. The tensile strength can be rewritten as (see \citealt{1997A&A...323..566L}):
\bea
S_{\max}&\simeq& 1.6\times 10^{6}(1-P)\left(\frac{\beta}{5}\right)\left(\frac{\bar{E}}{10^{-3}\rm eV}\right)\left(\frac{\alpha}{1}\right)\nonumber\\
&&\times \left(\frac{a}{5\rm nm}\right)^{-2}\left(\frac{0.3\rm nm}{h}\right) ~\erg\cm^{-3}.\label{eq:Smax_comp}
\ena

The tensile strength decreases rapidly with increasing the particle radius, as $a_{p}^{-2}$, and decreases with increasing the porosity $P$. In the following, we fix the porosity $P=0.2$, as previously assumed for {\it Planck} data modeling (\citealt{Guillet:2017hg}), and adopt the typical value of $\alpha=1$. 

\subsection{Silicate-core and amorphous carbon mantle grain model}
Finally, we consider a simple grain model including amorphous silicate core and carbonaceous material in the form of small grains or mantle (see Figure \ref{fig:grain_mod}, model c). Let $R_{1}$ be the radius of silicate core and $L$ be the thickness of the mantle. The effective grain size for this model is $a=R_{1}+L$. 

For the case of pure ice mantle, one can take the tensile strength of bulk ice $S_{\max}\sim 10^{7}\erg\cm^{-3}$ for the mantle layer. In the presence of small amorphous carbon grains, the tensile strength of the mantle layer might be increased considerably by several times (\citealt{Litwin:2012ii}).

The silicate core can be either compact or composite, but their nature is not important for the destruction of the mantle layer because even the bonds in the composite core are broken by the centrifugal force, the grain is only destroyed when the outer layer is ejected. Thus, we assume that the core is compact with $S_{\max}\sim 10^{9}-10^{10}\erg\cm^{-3}$.

\section{Grain Suprathermal Rotation by Radiative Torques}\label{sec:rot}


\subsection{Radiative torques of irregular grains}
Let $u_{\lambda}$ be the spectral energy density of radiation field at wavelength $\lambda$. The energy density of the radiation field is then $u_{\rad}=\int u_{\lambda}d\lambda$. To describe the strength of a radiation field, let define $U=u_{\rm rad}/u_{\rm ISRF}$ with 
$u_{\rm ISRF}=8.64\times 10^{-13}\erg\cm^{-3}$ being the energy density of the average interstellar radiation field (ISRF) in the solar neighborhoord as given by \cite{1983A&A...128..212M}. Thus, the typical value for the ISRF is $U=1$.

Radiative torque (RAT) arising from the interaction of an anisotropic radiation field with an irregular grain is defined as
\bea
{\Gamma}_{\lambda}=\pi a^{2}
\gamma_{\rm rad} u_{\lambda} \left(\frac{\lambda}{2\pi}\right){Q}_{\Gamma},\label{eq:GammaRAT}
\ena
where $\gamma_{\rm rad}$ is the anisotropy degree of the radiation field, ${Q}_{\Gamma}$ is the RAT efficiency, and $a$ is the effective size of the grain which is defined as the radius of the sphere with the same volume as the irregular grain (\citealt{1996ApJ...470..551D}; \citealt{2007MNRAS.378..910L}).

The magnitude of RAT efficiency, $Q_{\Gamma}$ can be approximated by a power-law (\citealt{Hoang:2008gb}):
\bea
Q_{\Gamma}\sim 0.4\left(\frac{{\lambda}}{1.8a}\right)^{\eta},\label{eq:QAMO}
\ena
where $\eta=0$ for $\lambda \lesssim 1.8a$  and $\eta=-3$ for $\lambda > 1.8a$. 

Numerical calculations of RATs for several shapes of different optical constants in \cite{2007MNRAS.378..910L} find the slight difference in RATs among the realization. An extensive study for a large number of irregular shapes by \cite{Herranen:2018wd} shows little difference in RATs for silicate, carbonaceous, and iron compositions. Moreover, the analytical formula (Equation \ref{eq:QAMO}) is also in a good agreement with their numerical calculations. Therefore, one can use Equation (\ref{eq:QAMO}) for the different grain compositions and grain shapes, and the difference is an order of unity

Let $\overline{\lambda}=\int \lambda u_{\lambda}d\lambda/u_{\rm rad}$ be the mean wavelength of the radiation field. For the ISRF, $\overline{\lambda}=1.2\mum$. The average radiative torque efficiency over the spectrum is defined as
\bea
\overline{Q}_{\Gamma} = \frac{\int \lambda Q_{\Gamma}u_{\lambda} d\lambda}{\int \lambda u_{\lambda} d\lambda}.
\ena

For interstellar grains with $a\lesssim \overline{\lambda}/1.8$, $\overline{Q}_{\Gamma}$ can be approximated to (\citealt{2014MNRAS.438..680H})
\bea
\overline{Q}_{\Gamma}\simeq 2\left(\frac{\overline{\lambda}}{a}\right)^{-2.7}\simeq 2.6\times 10^{-2}\left(\frac{\overline{\lambda}}{0.5\mum}\right)^{-2.7}a_{-5}^{2.7},
\ena
where $a_{-5}=a/10^{-5}\cm$, and $\overline{Q_{\Gamma}}\sim 0.4$ for $a> \overline{\lambda}/1.8$.

Therefore, the average radiative torque can be given by
\bea
\Gamma_{\rm RAT}&=&\pi a^{2}
\gamma_{\rm rad} u_{\rad} \left(\frac{\overline{\lambda}}{2\pi}\right)\overline{Q}_{\Gamma}\nonumber\\
&\simeq & 5.8\times 10^{-29}a_{-5}^{4.7}\gamma_{\rm rad}U\overline{\lambda}_{0.5}^{-1.7}\erg,~~~
\ena
for $a\lesssim \overline{\lambda}/1.8$, and
\bea
\Gamma_{\rm RAT}\simeq & 8.6\times 10^{-28}a_{-5}^{2}\gamma_{\rm rad}U\overline{\lambda}_{0.5}\erg,~~~
\ena
for $a> \bar{\lambda}/1.8$, where $\overline{\lambda}_{0.5}=\overline{\lambda}/0.5\mum$

The well-known damping process for a rotating grain is sticking collision with gas atoms, followed by evaporation. Thus, for a gas with He of $10\%$ abundance, the characteristic damping time is
\bea
\tau_{\gas}&=&\frac{3}{4\sqrt{\pi}}\frac{I}{1.2n_{\rm H}m_{\rm H}
v_{\rm th}a^{4}}\nonumber\\
&\simeq& 8.74\times 10^{4}a_{-5}\hat{\rho}\left(\frac{30\cm^{-3}}{n_{\H}}\right)\left(\frac{100\K}{T_{\gas}}\right)^{1/2}~{\rm yr},~~
\ena
where $\hat{\rho}=\rho/3\g\cm^{-3}$ with $\rho$ being the dust mass density, $v_{\rm th}=\left(2k_{\B}T_{\rm gas}/m_{\rm H}\right)^{1/2}$ is the thermal velocity of a gas atom of mass $m_{\rm H}$ in a plasma with temperature $T_{\gas}$ and density $n_{\H}$, the spherical grains are assumed (\citealt{2009ApJ...695.1457H}; \citealt{1996ApJ...470..551D}). This time is equal to the time required for the grain to collide with an amount of gas of the grain mass.

IR photons emitted by the grain carry away part of the grain's angular momentum, resulting in the damping of the grain rotation. For strong radiation fields or not very small sizes, grains can achieve equilibrium temperature, such that the IR damping coefficient (see \citealt{1998ApJ...508..157D}) can be calculated as
\bea
F_{\rm IR}\simeq \left(\frac{0.4U^{2/3}}{a_{-5}}\right)
\left(\frac{30 \cm^{-3}}{n_{\H}}\right)\left(\frac{100 \K}{T_{\gas}}\right)^{1/2}.\label{eq:FIR}
\ena 

Other rotational damping processes include plasma drag, ion collisions, and electric dipole emission. These processes are mostly important for PAHs and very small grains (\citealt{1998ApJ...508..157D}; \citealt{Hoang:2010jy}; \citealt{2011ApJ...741...87H}). Thus, the total rotational damping rate by gas collisions and IR emission can be written as
\bea
\tau_{\rm damp}^{-1}=\tau_{\gas}^{-1}(1+ F_{\rm IR}).\label{eq:taudamp}
\ena

For strong radiation fields of $U\gg 1$ and not very dense gas, one has $F_{\rm IR}\gg 1$. Therefore, $\tau_{\rm damp}\sim \tau_{\gas}/F_{\IR}\sim a_{-5}^{2}U^{2/3}$, which does not depend on the gas properties. In this case, the only damping process is IR emission.

For the radiation source with stable luminosity considered in this paper, radiative torques $\Gamma_{\rm RAT}$ is constant, and the grain velocity is steadily increased over time. The equilibrium rotation can be achieved at (see \citealt{2007MNRAS.378..910L}; \citealt{2009ApJ...695.1457H}; \citealt{2014MNRAS.438..680H}):
\bea
\omega_{\rm RAT}=\frac{\Gamma_{\rm RAT}\tau_{\rm damp}}{I},~~~~~\label{eq:omega_RAT0}
\ena
where $I=8\pi \rho a^{5}/15$ is the grain inertia moment.

\subsection{Strong radiation field}
For the case with $U\gg 1$, such as $F_{\rm IR}\gg 1$, plugging $\Gamma_{\rm RAT}$ (Equation \ref{eq:GammaRAT}) and $\tau_{\rm damp}$ (Equation \ref{eq:taudamp}) into the above equation, one obtain
\bea
\omega_{\rm RAT}&\simeq &7.1\times 10^{7}\gamma_{\rm rad,-1} a_{-5}^{1.7}U^{1/3}\bar{\lambda}_{0.5}^{-1.7}\rad\s^{-1},~~~\label{eq:omega_RAT}
\ena
for grains with $a\lesssim \bar{\lambda}/1.8$, and
\bea
\omega_{\rm RAT}&\simeq &\frac{1.1\times 10^{8}\gamma_{\rm rad,-1}}{a_{-5}}U^{1/3}\bar{\lambda}_{0.5}\rad\s^{-1},~~~
\ena
for grains with $a> \overline{\lambda}/1.8$.

\subsection{Weak radiation field}
In this case, both gas damping and IR emission is important. The rotation rate by RATs is given by
\bea
\omega_{\rm RAT}&\simeq &3.2\times 10^{7}\gamma_{\rm rad,-1} a_{-5}^{0.7}\bar{\lambda}_{0.5}^{-1.7}\nonumber\\
&\times&\left(\frac{U}{\hat{n}\hat{T}_{\rm gas}^{1/2}}\right)\left(\frac{1}{1+F_{\rm IR}}\right)\rad\s^{-1},~~~\label{eq:omega_RAT}
\ena
for grains with $a\lesssim \bar{\lambda}/1.8$, and
\bea
\omega_{\rm RAT}&\simeq &1.6\times 10^{8}\frac{\gamma_{\rm rad,-1}}{a_{-5}^{2}}\bar{\lambda}_{0.5}\nonumber\\
&&\times \left(\frac{U}{\hat{n}\hat{T}_{\rm gas}^{1/2}}\right)\left(\frac{1}{1+F_{\rm IR}}\right)\rad\s^{-1},~~~
\ena
for grains with $a> \overline{\lambda}/1.8$. Here $\gamma_{\rm rad,-1}=\gamma_{\rm rad}/0.1$ is the anisotropy of radiation field relative to the typical anisotropy of the diffuse interstellar radiation field.

\section{Maximum grain size constrained by Radiative Torque Disruption}\label{sec:disr}
In this section, we will quantify the effect of centrifugal force due to suprathermal rotation by RATs on grain properties. We consider a range of the radiation strength from $U\sim 10^{-3}-10^{3}$. The radiation anisotropy degree also varies with the location, between $\gamma_{\rm rad}\sim 0.1$ for the diffuse medium to $\gamma_{\rm rad}\sim 0.7$ for molecular clouds (\citealt{2007ApJ...663.1055B}), and $\gamma_{\rm rad}=1$ for grains close to a star. For numerical estimates below, we will assume an average value of $\gamma_{\rm rad}=0.5$, which is realistic for radiation on a cloud surface.

\subsection{Contact binary grain model}
The centrifugal force acting on the secondary grain of mass $M_{2}$ due to the rotation is 
\bea
F_{\rm Cen}=M_{2}r_{2}\omega^{2},\label{eq:Fcen}
\ena
where $r_{2}$ is the distance from the center of $M_{2}$ to the grain center of mass, as given by
\bea
r_{2}=\frac{M_{1}(R_{1}+R_{2})}{M_{1}+M_{2}}=\frac{\rho_{1}R_{1}(1+s)}{(\rho_{1}+\rho_{2}s^{3})},
\ena
with $r_{1}+r_{2}\approx R_{1}+R_{2}=R_{1}(1+s)$.

From Equation (\ref{eq:Fcen}) with Equation (\ref{eq:FJKR}), one can derive the critical rotation rate required to disrupt the binary grain as follows
\bea
\omega^{2}\ge\omega_{\rm disr}^{2}= \frac{9\gamma(\rho_{1}+\rho_{2}s^{3})}{4(1+s^{2})s^{2}\rho_{1}\rho_{2}}\frac{1+s^{3}}{a^{3}}.\label{eq:omega_disr2}
\ena

For $\rho_{1}\sim \rho_{2}=3\g\cm^{-3}$, one obtains
\bea
\omega_{\rm disr}\simeq 8.6\times 10^{7}\gamma_{1}\frac{(1+s^{3})}{s(1+s)}a_{-5}^{-3/2}~\rad\s^{-1}.
\ena

Using $\omega_{\rm RAT}$ from Equation (\ref{eq:omega_RAT}), one can calculate the the disruption size:
\bea
\left(\frac{a_{\rm disr}}{0.1\mum}\right)^{3.2}\simeq 1.2U^{-1/3}\bar{\lambda}_
{0.5}^{1.7} \frac{\gamma_{1}}{\gamma_{\rm rad,-1}}\frac{(1+s^{3})}{s(1+s)}
\ena
for strong radiation fields with $F_{\rm IR}\gg 1$, and
\bea
\left(\frac{a_{\rm disr}}{0.1\mum}\right)^{2.2}&\simeq& 2.6\bar{\lambda}_
{0.5}^{1.7} \frac{\gamma_{1}}{\gamma_{\rm rad,-1}}\frac{(1+s^{3})}{s(1+s)}\nonumber\\
&&\times(1+F_{\rm IR})\left(\frac{\hat{n}\hat{T}_{\rm gas}^{1/2}}{U}\right)
\ena
for arbitrary radiation fields. The disruption size $a_{\rm disr}$ is a maximum grain size $a_{\rm max}$ that suprathermally rotating grains can withstand the rotational disruption by RATs.

Table \ref{tab:adisr_bin} shows the disruption size for the different radiation strengths and gas density for a binary grain consisting of two identical spheres. Grains larger than $0.15\mum$ cannot be present in the form of contact binary grains, but smaller grains can be present in the form of mixed grains.

\begin{table}
\begin{center}
\caption{Maximum grain size for contact binary grain model}\label{tab:adisr_bin}
\begin{tabular}{l l l l l l} \hline\hline\\
\multicolumn{2}{l}{Gas density} & 
\multicolumn{2}{c}{$a_{\rm disr}(\mu m)$}\\
n$_{\H}(\cm^{-3})$ & U=0.1 & U=1 & U=10 & U$=10^{2}$ & U$=10^{3}$\cr
\hline\\
0.1 &  0.129 & 0.101 & 0.079 & 0.062 & 0.049\cr
1.0 &  0.147 & 0.103 & 0.079 & 0.062 & 0.049\cr
10 &  0.273 & 0.125 & 0.082 & 0.063 & 0.049\cr
30 &  0.436 & 0.167 & 0.089 & 0.063 & 0.049\cr
100 &  ND & 0.267 & 0.111 & 0.067 & 0.049\cr
1000 &  ND & ND & 0.264 & 0.102 & 0.055\cr
\cr
\hline
\multicolumn{5}{l}{{\it Notes}:~$R_{2}/R_{1}=1$.}\cr
\multicolumn{5}{l}{ND= No Disruption}\cr
\cr
\hline\hline
\end{tabular}
\end{center}
\end{table}

\begin{figure*}
\includegraphics[scale=0.5]{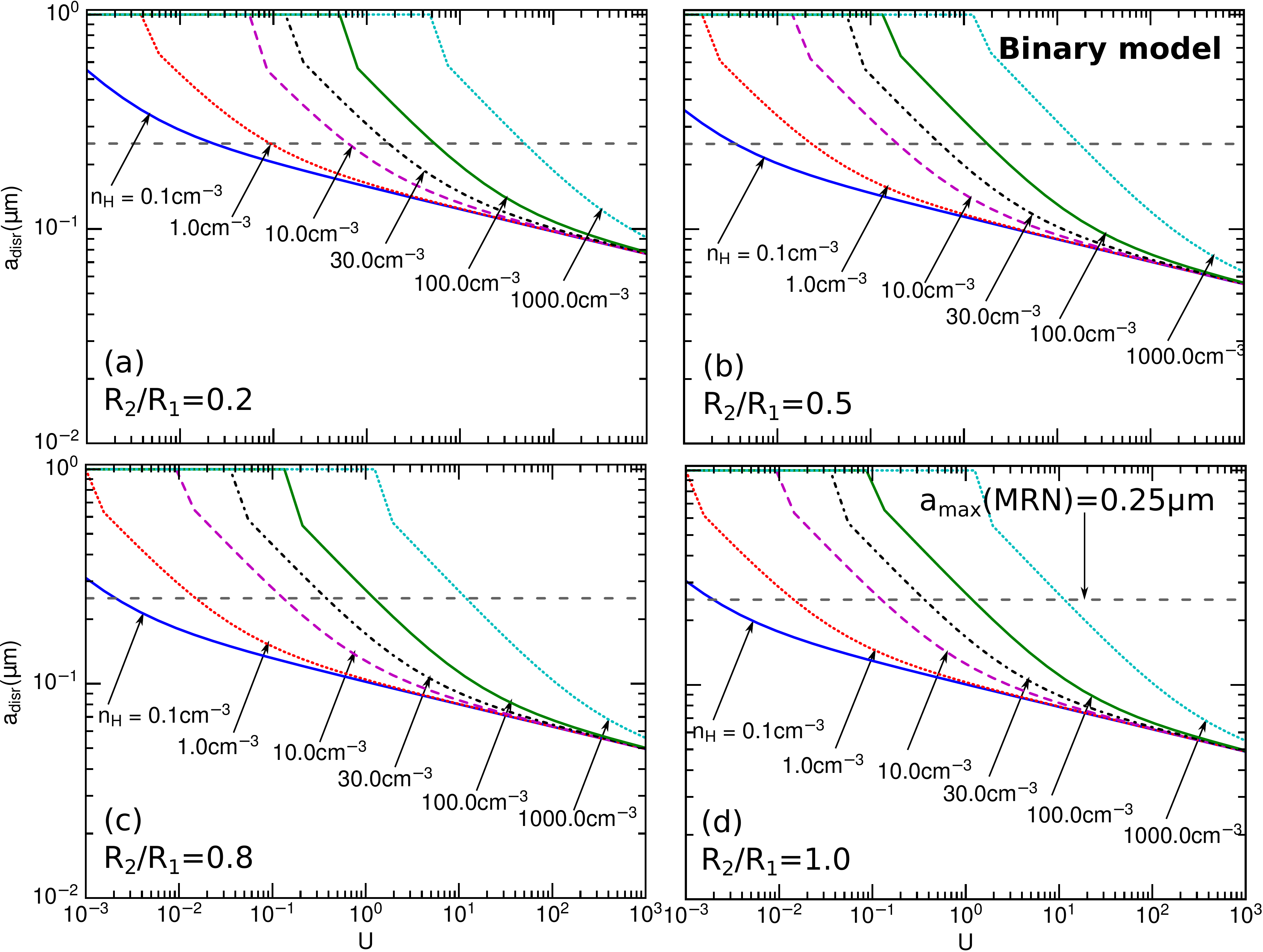}
\caption{Disruption grain size of a contact binary model as a function of the radiation strength for the different gas density $n_{\H}$, and we set $a_{\rm disr}=1.0\mum$ in case of no disruption. Four different values of the size ratio $s=R_{2}/R_{1}$ are considered. Horizontal dashed lines show the cutoff in MRN distribution of $a_{\max}=0.25\mum$. The transition occurs at $a_{\rm disr}\sim 0.67\mum$ due to the saturation of RATs for $a>\bar{\lambda}/1.8$.}
\label{fig:adisr_bin}
\end{figure*}

Figure \ref{fig:adisr_bin} shows the disruption grain size (also maximum size) as a function of the radiation strength $U$ for the different gas densities and various ratio of two spherical grains $s$. The disruption size decreases rapidly with increasing $U$, but it increases with increasing the gas density due to enhanced gas damping. At large $U$, the disruption size becomes weakly depends on the gas density due to the dominance of infrared emission damping. Note that in the case of no disruption, we set $a_{\rm disr}=1\mum$, which is likely the maximum grain size in the diffuse ISM. The issue of micron-sized grains is discussed in detail in Section \ref{sec:VLGs}.

\subsection{Composite grain model}
A spherical dust grain of radius $a$ rotating at velocity $\omega$ develops an average tensile stress due to centrifugal force which scales as (see \citealt{2018arXiv181005557H})
\bea
S=\frac{\rho a^{2} \omega^{2}}{4}.\label{eq:Stress}
\ena 

When the rotation rate is sufficiently high such as the tensile stress exceeds the maximum limit, namely tensile strength $S_{\rm max}$, the grain is disrupted. The critical rotational velocity is given by $S=S_{\rm max}$:
\bea
\omega_{\rm disr}&=&\frac{2}{a}\left(\frac{S_{\max}}{\rho} \right)^{1/2}\nonumber\\
&\simeq& \frac{3.6\times 10^{8}}{a_{-5}}S_{\max,7}^{1/2}\hat{\rho}^{-1/2}~\rad\s^{-1},\label{eq:omega_cri}
\ena
where $S_{\max,7}=S_{\max}/10^{7} \erg \cm^{-3}$ (\citealt{Hoang:2018es}).

For strong radiation fields such that $F_{\rm IR}\gg 1$, from Equations (\ref{eq:omega_RAT}) and (\ref{eq:omega_cri}), one can obtain the disruption grain size:
\bea
\left(\frac{a_{\rm disr}}{0.1\mum}\right)^{2.7}&\simeq&5.1\gamma_{\rad,-1}^{-1}U^{-1/3}\bar{\lambda}_
{0.5}^{1.7}S_{\max,7}^{1/2},~~~~~\label{eq:adisr_comp1}
\ena
for $a_{\rm disr}\le \overline{\lambda}/1.8$. 

For an arbitrary radiation field and $a\le \overline{\lambda}/1.8$., one obtains
\bea
\left(\frac{a_{\rm disr}}{0.1\mum}\right)^{1.7}&\simeq&11.4\gamma_{\rm rad,-1}^{-1}\bar{\lambda}_
{0.5}^{1.7}S_{\max,7}^{1/2}\nonumber\\
&&\times (1+F_{\rm IR})\left(\frac{\hat{n}\hat{T}_{\rm gas}^{1/2}}{U}\right),~~~\label{eq:adisr_comp2}
\ena
which depends on the local gas density and temperature due to gas damping.

Table \ref{tab:adisr_bin} shows the disruption size for the different radiation strengths and gas density, assuming the particle radius $a_{p}=5$ nm as in \cite{1989ApJ...341..808M}.

\begin{table}
\begin{center}
\caption{Maximum grain size for composite grain model}\label{tab:adisr_comp}
\begin{tabular}{l l l l l l} \hline\hline\\
\multicolumn{2}{l}{Gas density} & 
\multicolumn{2}{c}{$a_{\rm disr}(\mu m)$}\\
n$_{\H}(\cm^{-3})$ & U=0.1 & U=1 & U=10 & U$=10^{2}$ & U$=10^{3}$\cr
\hline\\
0.1 &  0.159 & 0.117 & 0.088 & 0.066 & 0.049\cr
1.0 &  0.190 & 0.121 & 0.088 & 0.066 & 0.049\cr
10 &  0.458 & 0.159 & 0.093 & 0.067 & 0.049\cr
30 &  ND & 0.239 & 0.104 & 0.068 & 0.050\cr
100 & ND & 0.450 & 0.139 & 0.072 & 0.051\cr
1000 &ND & ND & 0.4465 & 0.127 & 0.058\cr
\cr
\hline
\multicolumn{5}{l}{$^c$~$a_{p}=5$nm, $S_{\max}\sim 10^{6}\erg\cm^{-3}$}\cr
\cr
\hline\hline
\end{tabular}
\end{center}
\end{table}

\begin{figure*}
\includegraphics[scale=0.5]{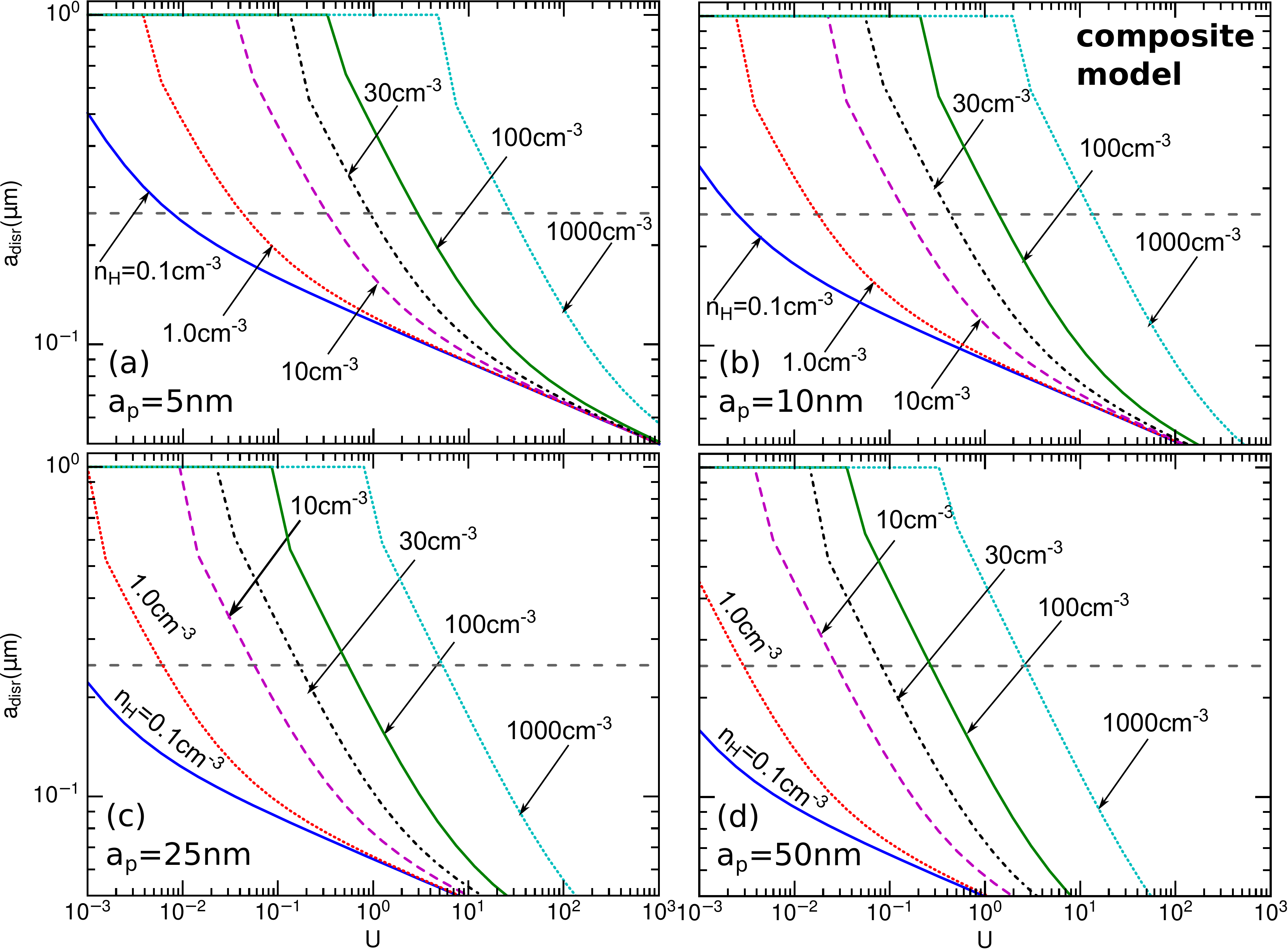}
\caption{Same as Figure \ref{fig:adisr_bin} but for a composite grain model with porosity $P=0.2$ and different particle radius $a_{p}$. The particle radius and its corresponding tensile strength are shown.}
\label{fig:adisr_comp}
\end{figure*}

Figure \ref{fig:adisr_comp} shows the disruption size as a function of $U$ for the different values of particle radius. The disruption size decreases significantly with increasing the particle radius $a_{p}$ due to the lower tensile strength. Increasing the gas density results in an increase in the disruption size due to the effect of gas collisional damping. For the typical ISM and $U=1$, the disruption size is $a_{\rm disr}\approx 0.1\mum$ for the particle radius $a_{p}\sim 25$ nm. 

\subsection{Core-mantle grain model}
We assume that the ice mantle is thick enough such that it can behaves like bulk ice. Therefore, the disruption of the core-mantle grain is not different from a compact grain, except the fact that in the later, the only mantle layer is ejected by the centrifugal force. The disruption size can be calculated by Equation (\ref{eq:adisr_comp1}) and (\ref{eq:adisr_comp2}) for $S_{\max}=10^{7}\erg\cm^{-3}$.

Table \ref{tab:adisr_mantle} shows the disruption size for core-mantle grains. Results for compact grains with $S_{\max}=10^{9}\erg\cm^{-3}$ are also shown for comparison.

\begin{table}
\begin{center}
\caption{Maximum grain size for core-mantle grain model}\label{tab:adisr_mantle}
\begin{tabular}{l l l l l l} \hline\hline\\
\multicolumn{2}{l}{Gas density} & 
\multicolumn{2}{c}{$a_{\rm disr}(\mu m)$ $^a$}\\
n$_{\H}(\cm^{-3})$ & U=0.1 & U=1 & U=10 & U$=10^{2}$ & U$=10^{3}$\cr
\hline\\
0.1 &  0.235 & 0.1719 & 0.129 & 0.097 & 0.073\cr
1.0 &  0.302 & 0.1801 & 0.129 & 0.097 & 0.073\cr
10 & ND & 0.262 & 0.139 & 0.098 & 0.073\cr
30 & ND & 0.422 & 0.162 & 0.101 & 0.073\cr
100& ND & ND & 0.237 & 0.110 & 0.074\cr
1000&ND & ND & ND & 0.223 & 0.089\cr
\cr
\hline
\multicolumn{5}{l}{$^a$~Results obtained for $S_{\max}=10^{7}\erg\cm^{-3}$}\cr
\cr
\hline\hline
\end{tabular}
\end{center}
\end{table}

\begin{figure*}
\includegraphics[scale=0.5]{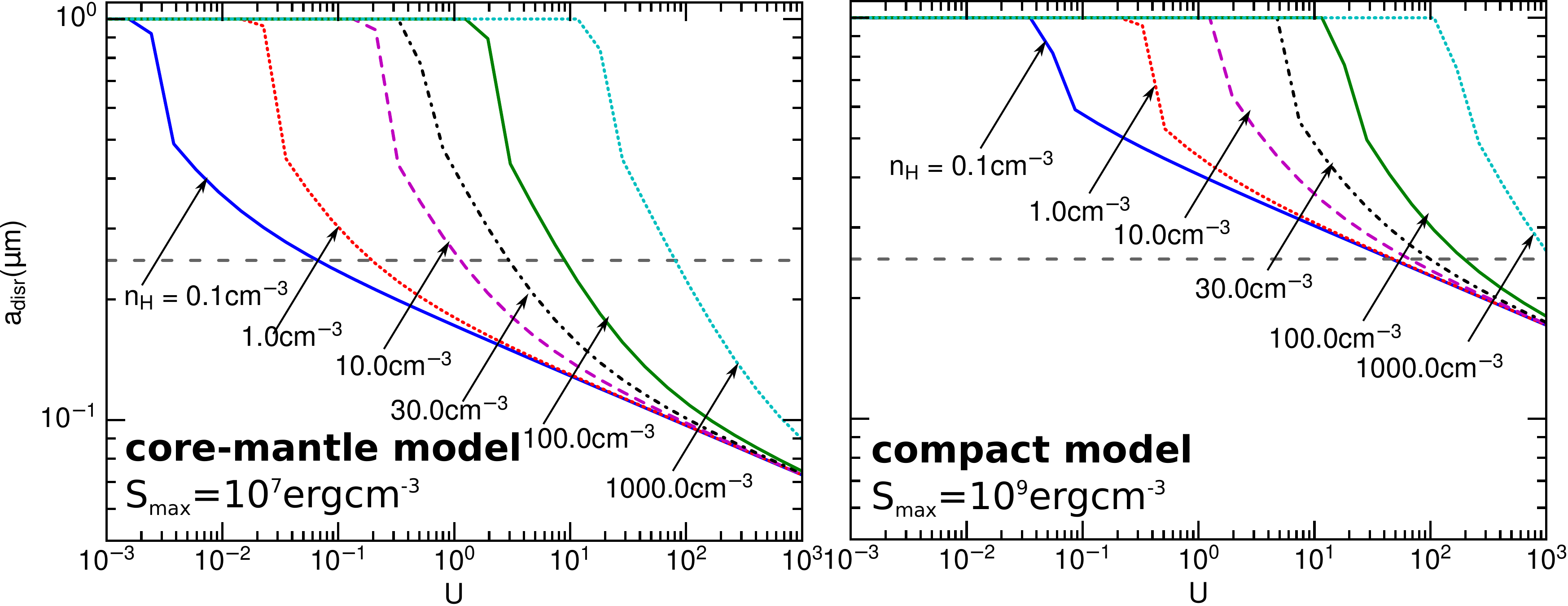}
\caption{Disruption size for core-mantle grain models with $S_{\max}=10^{7}\erg\cm^{-3}$ (left panel) and $10^{9}\erg\cm^{-3}$ (right panel).}
\label{fig:adisr_mantle}
\end{figure*}

Figure \ref{fig:adisr_mantle} shows the disruption size for the different gas density for a core-mantle grain (left panel) and compact grain (right panel). Same trend as Figures \ref{fig:adisr_bin} and \ref{fig:adisr_comp} are seen.

\section{Discussion}\label{sec:disc}
\subsection{Why do we need dynamical constraints for dust models?}
Interstellar dust models are usually constructed using observational constraints from starlight extinction and polarization (see \citealt{2003ARA&A..41..241D}). By varying the grain size distribution, a large number of dust models can successfully reproduce observational data (\citealt{2004ApJS..152..211Z}), including compact grains (\citealt{2001ApJ...548..296W}), composite (\citealt{1996ApJ...472..643M}), and core-mantle models (\citealt{1997A&A...323..566L}). As a consequence, one cannot get insight into internal structures of dust grains. In light of the discovery of rotational disruption effect by radiative torques \citep{2018arXiv181005557H}, in this paper, we showed that rotational disruption is an important dynamical constraint of the grain size distribution.

Table \ref{tab:adisr_ISM} shows the disruption size by RATD, which is the upper cutoff of the size distribution, for the different grain models, including binary, composite, and core-mantle models. Compact grains are not affected by the dynamical constraint if the radiation field is average or weak $U\le 1$, but all other models have the maximum size below $0.5\mum$. Composite grains made of tiny inclusions of $a_{p}\sim 5$ nm can survive in the ISRF if the grain size is below $0.25\mum$. For larger inclusions of $a_{p}\sim 25$ nm which have smaller tensile strength, large composite grains ($a>0.1\mum$) cannot survive because of disruption. As the radiation strength $U$ increases, the disruption size is decreased due to stronger RATD efficiency.

\begin{table}
\begin{center}
\caption{Maximum grain size for different dust models in the diffuse ISM}\label{tab:adisr_ISM}
\begin{tabular}{l l l l l l} \hline\hline\\
\multicolumn{2}{l}{Grain Model} & 
\multicolumn{2}{c}{$a_{\rm disr}(\mu m)$}\\
 & U=0.1 & U=1 & U=10 & U$=10^{2}$ & U$=10^{3}$\cr
\hline\\
Binary$^a$ &  0.523 & 0.198 & 0.102 & 0.072 & 0.055\cr
Binary$^b$ &  0.436 & 0.167 & 0.089 & 0.063 & 0.049\cr
Composite$^c$ &  ND & 0.239 & 0.103 & 0.068 & 0.050\cr
Composite$^d$ &  0.333 & 0.102 & 0.053 & 0.037 & 0.027\cr
Core-mantle$^e$ & ND& 0.421 & 0.162 & 0.100 & 0.073\cr
Compact$^f$ &  ND & ND & 0.484 & 0.249 & 0.174\cr

\cr
\hline
\multicolumn{5}{l}{{\it Notes}:~Diffuse ISM of $n_{\H}=30\cm^{-3}, T_{\gas}=100\K$.}\cr
\multicolumn{5}{l}{$^a$~$R_{2}/R_{1}=0.5$; $^b$~$R_{2}/R_{1}=1$}\cr
\multicolumn{5}{l}{$^c$~$a_{p}=5$ nm, $S_{\max}=1.2\times 10^{6}\erg\cm^{-3}$}\cr
\multicolumn{5}{l}{$^d$~$a_{p}=25$ nm, $S_{\max}=5.1\times 10^{4}\erg\cm^{-3}$}\cr
\multicolumn{5}{l}{$^e$~$S_{\max}=10^{7}\erg\cm^{-3}$}\cr
\multicolumn{5}{l}{$^f$~$S_{\max}=10^{9}\erg\cm^{-3}$}\cr
\cr
\hline\hline
\end{tabular}
\end{center}
\end{table}

One note that the RATD occurs on a timescale:
\bea
t_{\rm disr}&=&\frac{I\omega_{\rm disr}}{dJ/dt}=\frac{I\omega_{\rm disr}}{\Gamma_{\rm RAT}},\label{eq:tdisr}\\
&\simeq&10^{5}U^{-1}\overline{\lambda}_{0.5}^{1.7}S_{\max,7}^{1/2}\left(\frac{a_{\rm disr}}{0.1\mum}\right)^{-0.7}{~\rm yr},
\ena
which is much shorter than the shattering time by grain-grain collisions:
\bea
\tau_{\rm shat}&=&\frac{1}{\pi a^{2} n_{gr}v_{\rm gg}}=\frac{4\rho a M_{g/d}}{3n_{\H}m_{\H}v_{\rm gg}},\nonumber\\
&\simeq &2.5\times 10^{7}a_{-5}\left(\frac{30\cm^{-3}}{n_{\H}}\right)\left(\frac{1\rm km\s^{-1}}{v_{\rm gg}}\right) \rm yr,
\ena
where we assume the single size $a$ distribution with the gas-to-dust mass ratio $M_{g/d}= 100$ and the grain density $n_{\rm gr}$, and $v_{\rm gg}$ is the relative velocity of grains. This is much longer than the disruption time by RATD.

As a result, RATD can play an important role in constraining the maximum grain size of grains in the diffuse ISM, which is thought due to grain shattering \citep{Hirashita:2009p1139} induced by grain acceleration in magnetohydrodynamic (MHD) turbulence (\citealt{Yan:2004ko}; \citealt{Hoang:2012cx}). Because the RATD depends on the tensile strength of the grain, local gas properties and radiation field, the upper cutoff $a_{\max}$ is different for the different grain models and changes with the local conditions.

\subsection{Constraining grain internal structures with observations}
One of the mysterious issues of interstellar dust is the internal structure of dust grains, such as how constituents are organized. To date, no theoretical attempt have been made to relate the grain internal structure with observable quantities (i.e., emission and polarization). 

Here we suggest that the RATD effect can be used to constrain the internal structure because the RATD efficiency depends on the grain tensile strength which is characterized by the grain structure and compositions. The strategy is as follows. First, using observational constraints of extinction and polarization one can obtain the maximum grain size $a_{\max}$. By comparing $a_{\max}$ with the disruption size $a_{\rm disr}$, one then can constrain the tensile strength of grains. This can provide insight into whether grains are compact/core-mantle or composite. If the grain is composite, then, one can further infer the average radius of individual particles $a_{p}$ and porosity. 

The first potential scenario is to observe RATD in strong radiation fields such as supernovae and novae. Our estimates in \cite{2018arXiv181005557H} show the maximum grain size decreases with the decreasing the cloud distance to the supernova. If the cloud distance can be estimated such as through time evolution of color excess (\citealt{2018MNRAS.tmp.1547B}; \citealt{2018MNRAS.473.1918B}), one can then obtain the tensile strength by comparing the disruption size with the grain size estimated from reddening observations.

The second scenario is to use polarimetric observations. Indeed, the largest grains dominate the emission and polarization at long, far-infrared/submm wavelengths . Therefore, in the RAT alignment paradigm, the degree of far-IR/submm polarization first increases with increasing the radiation strength and then declines beyond some critical strength due to RATD. Interestingly, this trend might already be seen in Planck data \citep{Collaboration:2018ux}. Furthermore, as the radiation strength increases, the polarization curve of starlight becomes narrower and the peak wavelength shifts to short wavelengths, as a result of RAT alignment and reduction of largest grains by RATD.

\subsection{Are silicate and carbonaceous grains really separate?}
\cite{2006ApJ...651..268C} found that the 3.4 $\mum$ C-H feature is negligibly polarized, whereas the 9.7$\mum$ Si-O feature is strongly polarized for the light of sights toward the Galactic Center. The authors suggest that carbonaceous grains must be a separate component and these grains should be not aligned. Theoretically, if carbon grains are separate, they cannot be aligned due to their diamagnetic nature (\citealt{2016ApJ...831..159H}). The question now is why there are two separate silicate and carbonaceous materials?

Using our results in Section \ref{sec:disr}, we show that large composite grains ($a>0.1\mum$) with the particle radius $a_{p}>25$ nm are not stable in the ISM due to the low tensile strength, but composite grains with $a_{p}<10$ nm can survive RATD if their size is below $0.3\mum$ (see Table \ref{tab:adisr_ISM}). Therefore, if the original silicate and carbon grains are typically sized above $\sim$ 50 nm, then, the sticking collisions of these particles will form a composite grain, which is rapidly destroyed by RATD.

Moreover, carbon material can be mixed with silicate grains through contact binary or core-mantle models. For both models, large mixed grains can withstand the RATD. \cite{2002ApJ...577..789L} pointed out that the lack of 3.4 $\mum$ polarization is insufficient to reject core-mantle model.
A detailed study by \cite{Li:2014dk} shows that the polarization of $3.4\mum$ feature produced by the core-mantle model still exceeds the observational upper limit. Nevertheless, \cite{2013A&A...558A..62J} argued that if the mantle is much thinner than the core radius, then, the $3.4\mum$ feature is absent from core-mantle grain, which can explain the negligible polarization of $3.4\mum$ feature (see \citealt{2016RSOS....3p0224J} for a review). 

An alternative explanation is that the average radiation along the line of sight toward GC is enhanced such that RATD can disrupt mixed grains, including composite and core-mantle ones.

Finally, one note that in molecular clouds, RATD is inefficient due to weak radiation fields and high gas density. As a result, large mixed silicate-carbonaceous grains can be present. This prediction would be tested with polarimetric observations. 

\subsection{On the evidence of micron-sized grains in the ISM}\label{sec:VLGs}

Interstellar dust is widely believed to comprise sub-micron sized-grains. Yet, numerous observations reveal the flat mid-IR ($\lambda\sim 3-8\mum$) extinction toward diffuse, translucent, and dense clouds, which can mostly be reproduced with micron-sized grains (hereafter very large grains-VLGs; \citealt{2013ApJ...773...30W}; \citealt{2014P&SS..100...32W}). Moreover, in-situ observations by spacecraft (\citealt{1994A&A...286..915G}; \citealt{Westphal:2014gf}) report the presence of VLGs in the interplanetary medium.
 
In the RATD picture, the presence of VLGs in translucent and dense clouds is not surprising because the RATD mechanism is inefficient due to high gas density and low radiation strength. For the diffuse ISM, VLGs could be mostly present in the environments with radiation intensity lower or local gas density is higher than the average ISM. For example, the combination of a slightly dense regions of $n\sim 100\cm^{-3}$ and $U\lesssim 0.1$ can increase the disruption size to $a_{\rm disr}\sim 5\mum$ using Equation (\ref{eq:adisr_comp2}) with $a_{p}=5$ nm.

Second, the existence of VLGs in the diffuse ISM reveals that such grains likely have compact structures with a high tensile strength (e.g., $S_{\rm max}\gtrsim 10^{9}\erg\cm^{3}$) which are not disrupted by RATD with average radiation fields (see Figure \ref{fig:adisr_comp}). Incidentally, this idea supports the results obtained by \cite{2015ApJ...811...38W} in which including a graphite component of micron sizes can successfully reproduce the observed mid-IR extinction.

Third, VLGs having a core-thick ice mantle structure can also survive against RATD. Indeed, following \cite{Hoang:2019td}, the maximum size of grains with ice mantles that can still be disrupted by RATD is given by
\bea
a_{\rm disr,max}&\simeq& 0.96\gamma_{\rm rad}\bar{\lambda}_{0.5}\left(\frac{U}{\hat{n}\hat{T}_{\gas}^{1/2}}\right)^{1/2}\left(\frac{1}{1+F_{\rm IR}}\right)\nonumber\\
&&\times \rho_{\rm ice}S_{\max,7}^{-1/2}~\mum,\label{eq:adisr_up}
\ena 
where $\rho_{\rm ice}\approx 1\g\cm^{-3}$ is the ice mass density. The equation gives $a_{\rm disr,max}\sim 1.2\mum$ for the tensile strength of bulk ice of $S_{\max}\approx 10^{7}\erg\cm^{-3}$ and $\gamma_{\rm rad}=0.5$. Therefore, micron-size ice grains are not disrupted by RATD. 

The presence of such micron-sized water ice grains in the diffuse ISM can resolve the crisis of interstellar oxygen as suggested in previous works (\citealt{2009ApJ...700.1299J}; \citealt{2015ApJ...801..110P}; \citealt{2015MNRAS.454..569W}).

Finally, if VLGs have composite fluffy structures, then, the contribution of chemical bonds is expected to increase the mean intermolecular energy from van der Waals, leading to $\alpha\gg 1$. Thus, the tensile strength is increased from the typical value shown in Equation (\ref{eq:Smax_comp}). Experimental measurements in \cite{Gundlach:2018cu} show that the tensile strength of composite grains with the constituent particles of radius $a_{p}\sim 2.4-0.15\mum$ can be fitted as $S_{\rm exp}\simeq 2.73\times 10^{5}(0.1\mum/a_{p})$. Therefore, comparing $S_{\exp}$ with Equation (\ref{eq:Smax_comp}), one obtains $\alpha\sim 4, 21$ for $a_{p}=5, 25$ nm, respectively. The increase of $S_{\max}$ ($\alpha$) raises the disruption size, $a_{\rm disr}$, by $21^{1/5.4}\sim 1.8$ (see Equation \ref{eq:adisr_comp1}). Moreover, from Equation \ref{eq:adisr_up}, one derives the maximum disruption size of composite VLGs as $a_{\rm disr,max}\sim 109(S_{\max}/10^{4}\erg\cm^{-3})^{-1/2}\mum$. Thus, From Table \ref{tab:adisr_ISM}, one can see that composite VLGs of $a\sim [a_{\rm disr}-a_{\rm disr, max}]$ can still be destroyed by RATD using the measured tensile strength.

\section{Summary}\label{sec:summ}
Using the RATD effect discovered by \cite{2018arXiv181005557H}, we have introduced a new dynamical constraint for interstellar dust models and studied implications of this constraint. The main results are summarized as follows:

\begin{itemize}

\item{} For all dust models except compact grains, we find that large grains of size $a>0.45\mum$ are destroyed by RATD in the average ISRF (i.e., $U=1$). Stronger radiation fields result in the disruption of smaller grains.

\item{} For the composite model, we find that large composite grains made of small individual particles of radius $a_{p}\le 25$nm can survive in the average ISRF with the upper limit of $a_{\max}\sim 0.24\mum$, which is incidentally similar to the upper cutoff of MRN distribution. The maximum size decreases to $a_{\max}\sim 0.1\mum$ for $U=10$. As a result, large composite grains can survive in the diffuse ISM with an average radiation field of $U<10$.
 
\item{} The growth toward micron-sized grains of non-compact structures in the diffuse ISM would be prohibited by RATD, but it would proceed in weak radiation fields such as dense molecular clouds. 
 
\item{} We explain the non-detection of polarization at the 3.4$\mum$ C-H feature by means of two separate silicate and carbonaceous dust materials which are disrupted by RATs from original composite grains made of large individual particles of radius $a_{p}\ge 50$ nm.

\item{} Using the RATD effect, we suggest that internal structures of grains can be constrained by observations of starlight extinction and polarization for the conditions with varying radiation fields, such as in the vicinity of a star, supernova, and nova.

\item{}We suggest that VLGs as required to reproduced mid-IR extinction likely have either compact structures of high tensile strength or a core-thick ice mantle, or they are located in regions with higher gas density and lower radiation strength than the average ISM.

\end{itemize}

\acknowledgments
We are grateful to the anonymous referee for useful comments and suggestions. We thank B.T Draine, V. Guillet, A. Lazarian, and P. Lesaffre for discussions on various issues related to rotational disruption and its implication. This work was supported by the Basic Science Research Program through the National Research Foundation of Korea (NRF), funded by the Ministry of Education (2017R1D1A1B03035359).


\bibliography{ms.bbl}

\end{document}